\documentclass[twocolumn,showpacs,preprintnumbers,amsmath,amssymb]{revtex4}

\usepackage{graphicx}
\usepackage{dcolumn}
\usepackage{bm}

\begin{document}

\title{Noise invoked resonances near a homoclinic bifurcation \\ in the glow discharge plasma}%
\author{Md. Nurujjaman}
\email{md.nurujjaman@saha.ac.in}
\affiliation{Plasma Physics Division,
Saha Institute of Nuclear Physics,
1/AF, Bidhannagar,
Kolkata -700064, India.}

\author{P. Parmananda}
\affiliation{Facultad de Ciencias, UAEM, Avenida Universidad 1001,
Colonia Chamilpa, Cuernavaca, Morelos, M\'{e}xico,}
\affiliation{Plasma Physics Division, Saha Institute of Nuclear Physics,
1/AF, Bidhannagar, Kolkata -700064, India}
\affiliation{Department of Physics, Indian Institute of Technology Bombay, Powai,
Mumbai 400076, India}
\author{A.N. Sekar Iyengar}
\email{ansekar.iyengar@saha.ac.in}
\affiliation{Plasma Physics Division, Saha Institute of Nuclear Physics,
1/AF, Bidhannagar, Kolkata -700064, India.}

\begin{abstract}
Stochastic Resonance (SR) and Coherence Resonance (CR)
have been  studied experimentally in the discharge plasma close
to a homoclinic bifurcation. For the SR phenomena, it is
observed that a superimposed subthreshold periodic signal can be
recovered via stochastic modulations of the discharge voltage.
Furthermore, it is realized that even in the absence of a
subthreshold deterministic signal, the system dynamics can be
recovered and optimized using noise. This effect is defined as CR
in the literature. In the present experiments,  induction
of SR and CR are  quantified using  the Absolute Mean Difference (AMD)
and Normalized Variance (NV) techniques respectively. AMD is a new statistical tool to 
quantify regularity in the stochastic resonance and is independent of lag.
\end{abstract}
\maketitle

\section{Introduction}
Stochastic Resonance (SR) is a phenomena in which the response of the
nonlinear system to a weak periodic input signal is amplified/optimized by
the presence of a particular level of noise~\cite{JPhysA:benzi}, i.e,
a previously untraceable subthreshold signal applied to a nonlinear system,
can be detected in the presence of noise. Furthermore, there
exists an optimal level of noise for which the  most efficient detection
takes place~\cite{prl:bruce,pre:parmananda}.
SR has been observed in many physical, chemical and
biological systems ~\cite{prl:bruce,pre:parmananda,JStatPhys:moss,prl:longtin,
JPhysChem:foster,JPhysChem:Amemiya,pre:santos,prl:kitajo}.
Coherence Resonance (CR) is the  phenomena wherein
regularity of the  dynamical behavior emerges  by virtue of
an interplay between the autonomous nonlinear dynamics
and the superimposed stochastic fluctuations. In CR, analogous
to  SR,  the extent of provoked  regularity
depends upon the amplitude of added
noise~\cite{prl:gang,prl:pikovsky}. The CR effect too  has been
studied exhaustively, both theoretically and experimentally, in
a wide range of nonlinear
systems~\cite{prl:Lin I,pop:dinklage,prl:Giacomelli,pre:miyakawa,pre:Santos1,prl:avila}.

The nonlinearity in plasma systems arises  from the most
fundamental processes, namely the wave-wave and wave-particle interactions.
Different modes may be excited due to nonlinear coupling
of waves and plasma components and the character of the
oscillations is  primarily determined by the plasma parameters and
perturbations~\cite{pr:duncan,pr:sturrock,pop:Shokri}.
In the present work, possibility of observing noise
invoked resonances in a glow discharge plasma is explored.
However, as a precursor to the experiments involving
noise, a systematic analysis of the autonomous dynamics
is performed. This includes  identification and
characterization of the bifurcation   in the
vicinity of the set-point employed for the
noise related experiments.
\section{Experimental setup}
\label{section:Experimental setup }
\begin{center}
\begin{figure}[ht]
\includegraphics[width=8.5 cm]{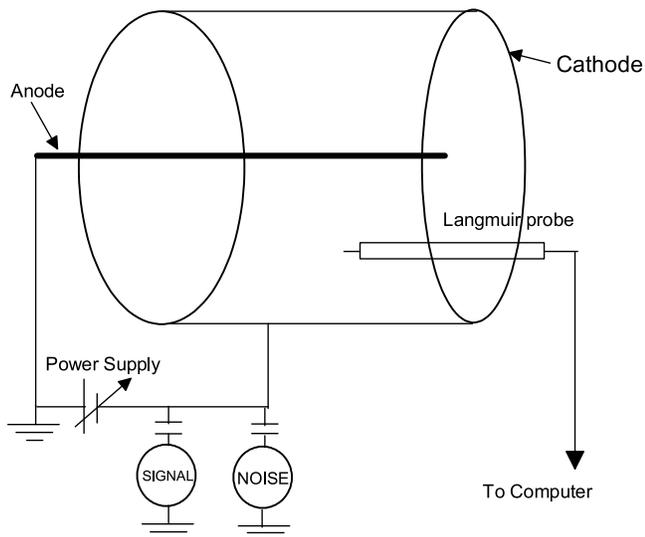}
\caption{Schematic diagram of the cylindrical
electrode system of the glow discharge plasma.
The probe was placed at a distance $l \approx12.5$ mm
from the anode. Signal and noise sources were
coupled to the Discharge Voltage (DV) through a capacitor.}
\label{fig1:setup}
\end{figure}
\end{center}

The experiments were performed in a hollow cathode
dc glow discharge plasma. The schematic diagram of
the experimental setup  is presented in Fig~\ref{fig1:setup}. A
hollow Stainless Steel (S.S) tube of length $\approx7~cm$ and of
diameter ($\phi$) $\approx$ 45 mm was used as the cathode and
a central rod of length $\approx7~cm$ and $\phi\approx$ 1.6 mm was employed
as the anode. The whole assembly was mounted inside a
vacuum chamber and was pumped down to a pressure
of  about  0.001 mbar using a rotary
pump. The chamber was subsequently filled with the
Argon gas up to a pre-determined  value of neutral
pressure by a needle valve. Finally  a discharge was struck
by a dc Discharge Voltage (DV), which could be varied in
the range of 0$-$1000 V.
The noise and subthreshold periodic square pulse generators were
coupled with DV through a capacitor [Fig.~\ref{fig1:setup}]. In
all the experiments DV was used as  the bifurcation parameter
while the remaining system parameters like pressure etc.,
were maintained constant. The system observable
was the electrostatic floating potential, which
was measured using a Langmuir probe of
diameter $\phi$ = 0.5 mm  and  length
2 mm. The tip of this Langmuir probe was placed in
the center of the electrode system as indicated in
Fig.~\ref{fig1:setup}.
The  plasma density and the electron
temperature were determined to be of the
order of 10$^7$cm$^{-3}$ and 3$-$4 eV respectively.
Furthermore, the  electron plasma frequency ($f_{pe}$)
was observed to be around  28 MHz, whereas the ion plasma
frequency ($f_{pi}$) was measured to be around  105 kHz.
\section{Autonomous Dynamics}
\label{section:Autonomous Dynamics}
\begin{center}
\begin{figure}[ht]
\includegraphics[width=8.5 cm]{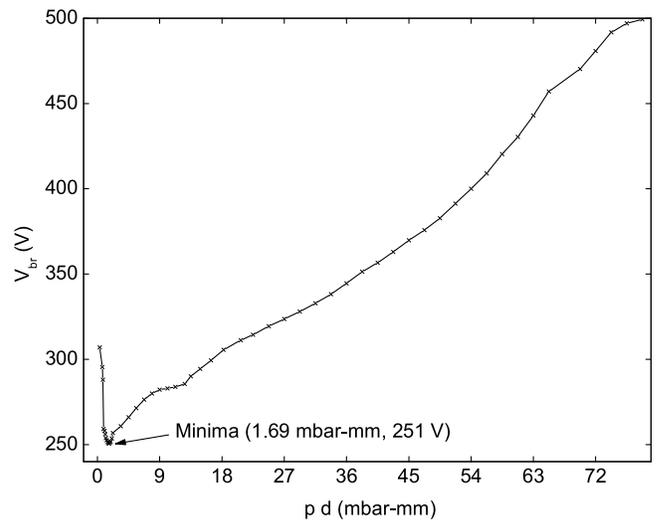}
\caption{ $V_{br}$ vs $pd$ (Paschen curve) for our experimental system. The
minimum occurs at (1.69 mbar-mm, 251 V). The system is excitable for
the $pd>$ minimum of the curve.}
\label{fig:Paschen}
\end{figure}
\end{center}

\begin{figure}[ht]
\includegraphics[width=8.5 cm]{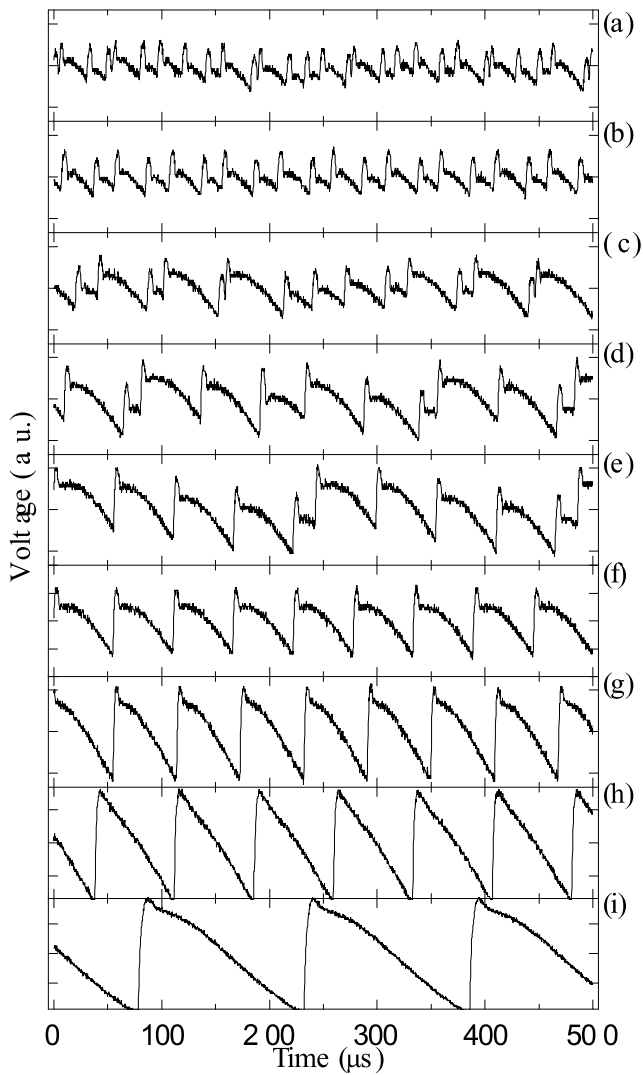}
\caption{Sequential change in the raw signal (normalized) at 0.89 mbar for different voltages: (a) 288 V; (b) 291 V; (c) 295 V; (d) 301 V; (e) 304 V; (f) 307 V; (g) 327 V; (h) 385 V; (i) 466 V. All y-axes range form -1 to 1.}
\label{fig:raw0.89mb}
\end{figure}

Before studying the noise induced dynamics, we characterized
the behavior of the autonomous system. Not surprisingly,
it was observed that at  different chamber pressures,
discharge struck at different voltages. Fig~\ref{fig:Paschen}
shows the breakdown voltage ($V_{br}$) at different $pd$,
where p and d are the filling pressure and radius of the
cathode respectively. This breakdown voltage ($V_{br}$) initially
decreases with an increase in $pd$, goes through a
minimum value resembling  a typical Paschen curve
and then begins to increase with increasing $pd$.
It is observed that the system is excitable for
the region $pd>$ Paschen minima. In the lower side it shows Self Organized Criticality~\cite{Physleta:jaman}. In this excitable
domain, the system dynamics are irregular (complex) at
the initial stages of the discharge voltage and upon
increasing  DV they  become regular (period-one) as shown in Fig~\ref{fig:raw0.89mb}.
Further augmentation of DV modifies  the  oscillation profile
and  results in the induction of typical relaxation
oscillations~\cite{chaos:Nurujjaman}.

\begin{center}
\begin{figure*}[ht]
\includegraphics{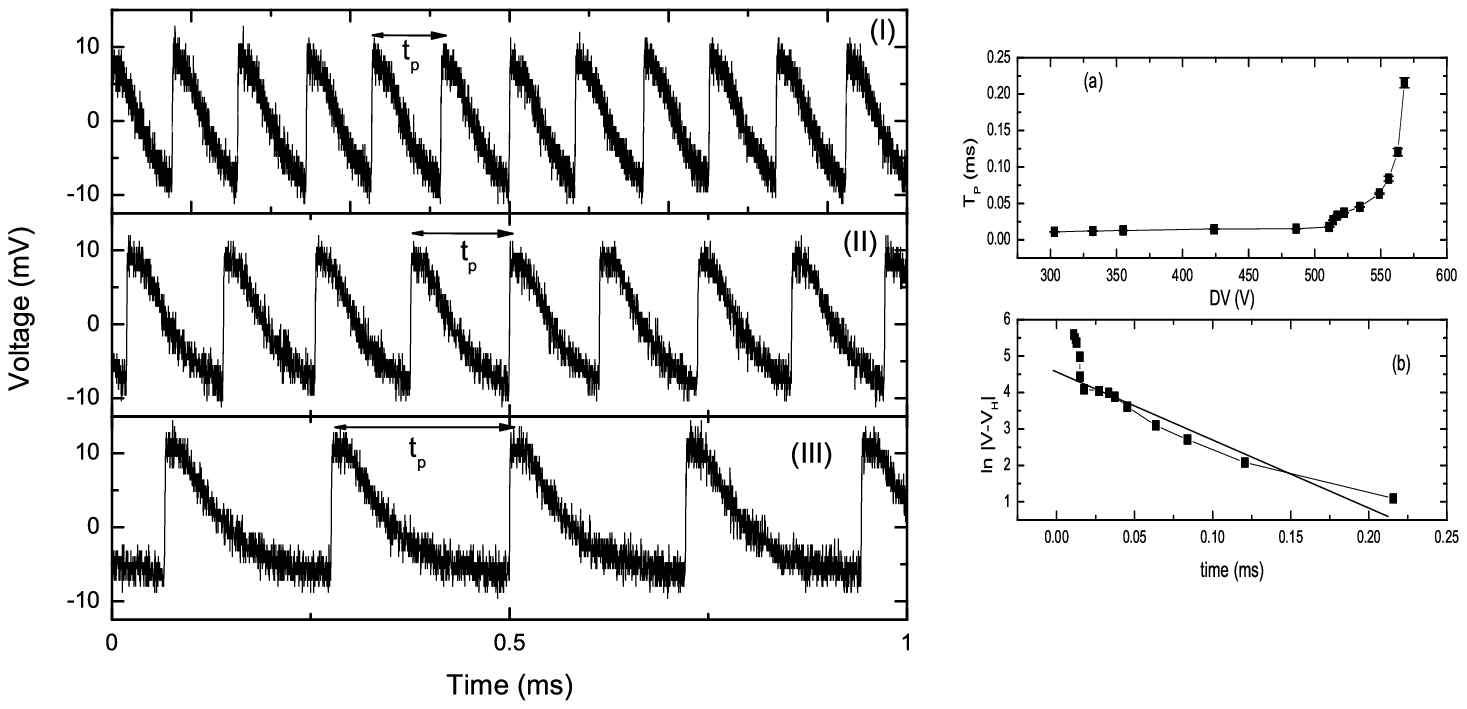}
\caption{Left panel: Timeseries showing that the period of
relaxation oscillations increases with
augmenting DV. Right panel:(a) Exponential increment of the
time period (T) with DV and (b) $\ln|V-V_H|$ vs T curve can
be fitted by a straight line indicating an underlying
homoclinic bifurcation.}
\label{fig:homoclin}
\end{figure*}
\end{center}

The time period (T) of these relaxation oscillations
increases  dramatically upon further incrementing
DV. This eventually results in the vanishing of the limit
cycle behavior beyond a critical DV ($V_H$). For larger
values of DV, the autonomous dynamics exhibit a steady state
fixed point behavior. Time traces from top to bottom in the
left panel of  Fig~\ref{fig:homoclin} depict this period
lengthening of the  oscillatory behavior. A systematic  analysis of
the increment in the period (T), presented in
Fig.~\ref{fig:homoclin}[(a) right panel], indicates
that the autonomous dynamics undergo a critical (exponential)
slowing down. Consequently, the  $\ln|V-V_H|$ vs T
curve can be fitted by a straight line, where $V_H$ is the
bifurcation point separating the oscillatory domain and the
steady state behavior. The results of Fig~\ref{fig:homoclin}
indicate that the system dynamics undergo a homoclinic
bifurcation at $V_H$ resulting in the loss of oscillations.

\begin{center}
\begin{figure*}[ht]
\includegraphics[]{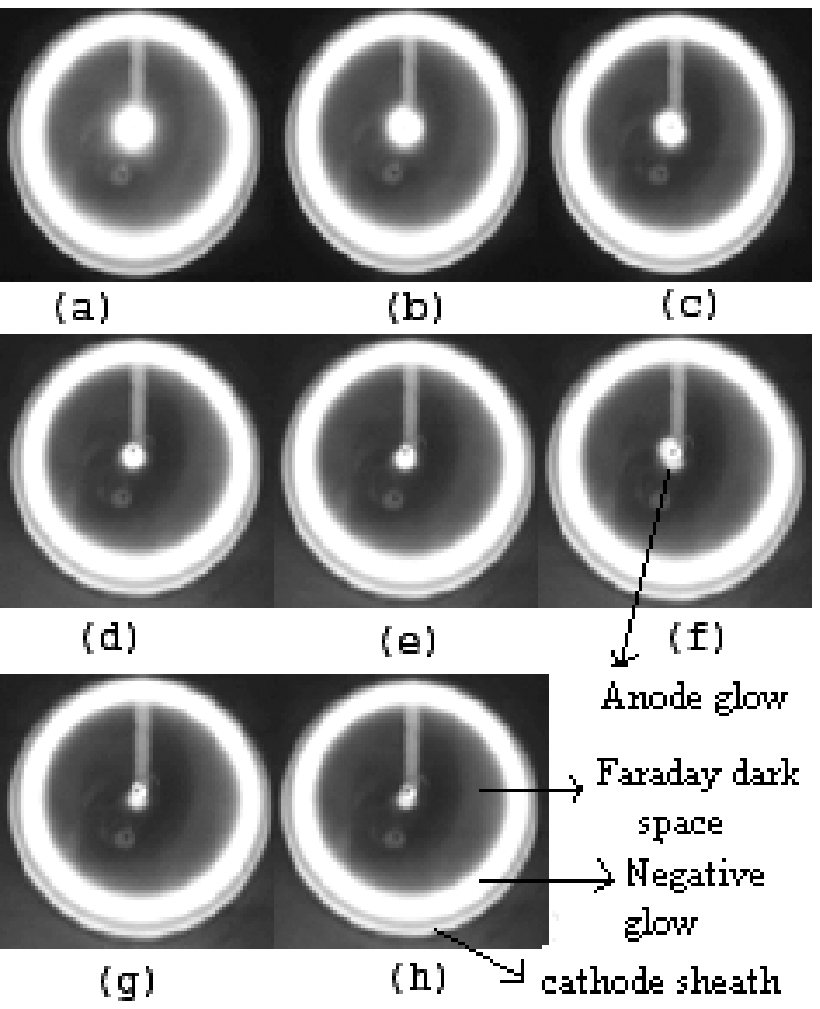}
\caption{Evolution of glow size of the anode glow with increasing DV (a) - (h).}
\label{fig:glow}
\end{figure*}
\end{center}
An anode glow is observed with these oscillations. 
Figs ~\ref{fig:glow}(a) shows that the glow with largest size, appears when the discharge is struck at a typical pressure of 0.95 mbar and its size decreases with increase in the DV until it finally disappears [Figs~\ref{fig:glow}(a)$-$\ref{fig:glow}(h)]. This may some types of unstable structure in the plasma and produces such oscillation of the instabilities~\cite{chaos:Nurujjaman}.
\begin{center}
\begin{figure*}[ht]
\includegraphics{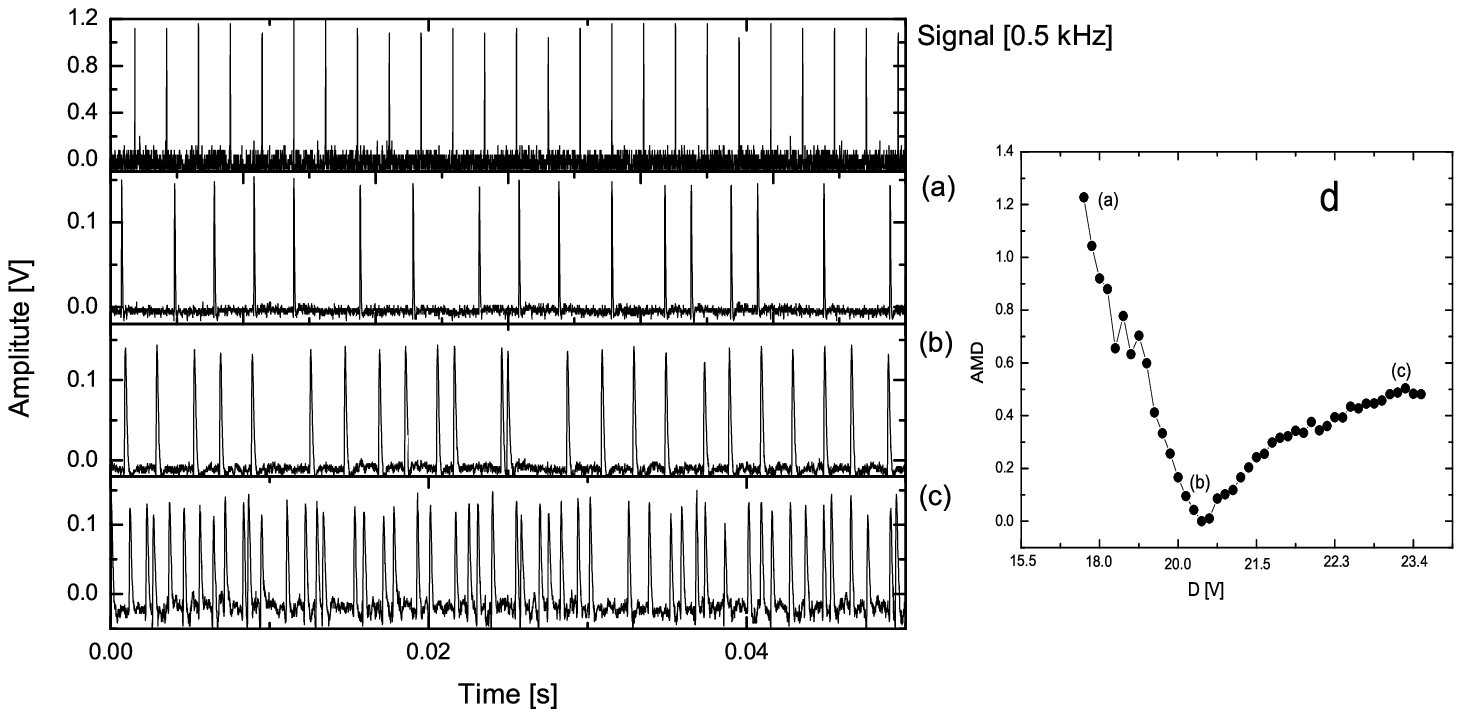}
\caption{Noise induced system response in the floating potential
fluctuations for low, medium , and high amplitude noise in
conjunction with a subthreshold periodic square pulse. The right
panel [Fig~\ref{fig:periodic}] shows the AMD as a function of
noise amplitude for the experiment performed at $V_0=307$ V and
pressure= 0.39 mbar. Left panel shows the subthreshold periodic
pulse train and the  three time series of floating
potential fluctuations at low level noise (a); at optimum
noise value (b) and at high amplitude noise (c).}
\label{fig:periodic}
\end{figure*}
\end{center}

\begin{center}
\begin{figure*}[ht]
\includegraphics{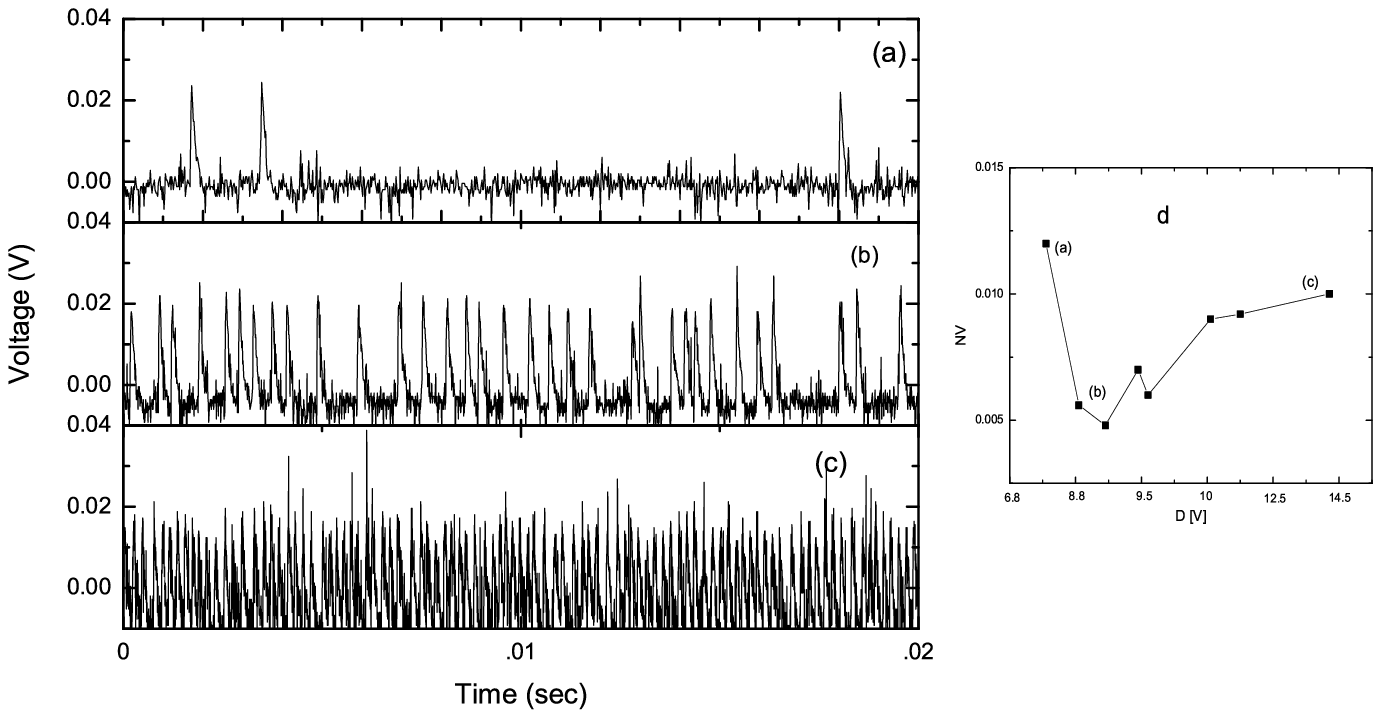}
\caption{Emergence of coherence resonance: The right
panel  shows the NV as a function
of noise amplitude for the experiments performed
at $V_0=344$ V and pressure = 0.5 mbar. Left panel: The
time series of  the floating potential fluctuations
for (a) low noise (b) optimum noise and (c) high level noise}
\label{fig:NV}
\end{figure*}
\end{center}

\section{Noise induced resonance in glow discharge}
In this section experimental results involving noise
generated resonances, namely  SR and CR are presented.
\subsection{Stochastic Resonance}
\label{subsection:SR}
For our experiments on stochastic resonance, the reference
voltage $V_0$ was chosen such that $V_0>V_H$ and therefore
the autonomous dynamics, by virtue of
an underlying homoclinic bifurcation, exhibit
steady state behavior. The discharge voltage $V$
was thereafter perturbed  $V=V_0+S(t)+D\xi$,
where  $S(t)$ is the subthreshold periodic pulse train
chosen for which $V=V_0+S(t)>V_H$, (subthreshold
signal does not  cause the  system to cross over
to the oscillatory regime) and $D\xi$
is  the added Gaussian white noise $\xi$ with
amplitude  $D$. Subthreshold periodic square pulse of
width $20~\mu s$ and duration 2 ms was constructed
using Fluke PM5138A function generator. Meanwhile,  the
gaussian noise  produced using the HP 33120A noise
generator was subsequently amplified using a noise
amplifier.

Fig.~\ref{fig:periodic}$(a)-(c)$ show time
series of the system response  in the
presence of an identical subthreshold
signal for three different
amplitudes of imposed noise. The subthreshold
periodic pulse train is also plotted, in the
top most graph of the left panel, for comparison
purposes. Fig.~\ref{fig:periodic}(a) shows
that there is little correspondence between the subthreshold
signal and the system response for a low noise amplitude.
However, there is excellent correspondence at an intermediate noise
amplitude [Fig.~\ref{fig:periodic}(b)]. Finally,  at higher
amplitudes of noise the subthreshold signal is lost amidst
stochastic fluctuations of the system
response [Fig.~\ref{fig:periodic}(c)]. Absolute
mean difference (AMD),  used to quantify the
information transfer between the subthreshold
signal and the system response, is defined
as $AMD=abs(mean(\frac{t_p}{\delta}-1))$.
$t_p$ and $\delta$ are the inter-peak
interval of the response signal
and mean peak interval of the subthreshold periodic
signal respectively. 

Fig~\ref{fig:periodic}(d) shows
that the experimentally computed AMD versus noise amplitude D curve
has a unimodal structure typical for the SR phenomena.
The minima in this curve  corresponds to
the optimal noise level for which maximum information transfer
between the input and the output takes place.
\subsection{Coherence Resonance}
\label{subsection:CR}
For the experiments on coherence resonance
DV $(V_0$) was located such that
the floating potential fluctuations
exhibit  fixed point behavior.
In order to minimize the effect of parameter drift,
a set-point ($V_0$) quite far from the homoclinic
bifurcation ($V_H$) was chosen. Subsequently, superimposed
noise on the discharge voltage was increased and the
provoked dynamics analyzed. The normalized variance (NV) was
used to quantify the extent of induced
regularity. It is defined as $NV=std(t_p) / mean(t_p)$,
where $t_p$ is the time elapsed between successive peaks. It
is evident that  more regular the induced
dynamics the lower the value of the computed NV.
For purely periodic dynamics the NV goes to zero.

Fig.~\ref{fig:NV}$(a)-(c)$ (left panel)
show the time series of the floating potential fluctuations
for different noise levels and Fig~\ref{fig:NV}(d) (right panel)
is the experimental NV curve as a function of noise
amplitude D. The point (a) in Fig~\ref{fig:NV}(d)
(time series shown in Fig.~\ref{fig:NV}(a)) is associated
with a low level of noise where the activation
threshold is seldom crossed, generating a sparsely
populated irregular spike sequence. As the noise
amplitude is increased, the NV decreases,
reaching a minimum (b) in Fig~\ref{fig:NV}(d)
(time series shown in Fig.~\ref{fig:NV}(b))
corresponding to an optimum noise level where maximum regularity of
the  generated spike sequence is observed.  As the
amplitude of superimposed noise is increased
further, the observed regularity is destroyed
manifested by an increase in the
NV; label (c) in Fig~\ref{fig:NV}(d)
(time series shown in Fig.~\ref{fig:NV}(c)).
This is a consequence of the dynamics being dominated by noise.
\section{Discussion}
The effect of noise has been studied experimentally near
a homoclinic bifurcation in glow discharge plasma
system. Our study demonstrates the  emergence of SR for
periodic subthreshold square pulse signals and the
induction of CR via purely stochastic fluctuations.
In SR experiments, the efficiency of information transfer
was quantified using AMD instead of the  power
norm which has been utilized elsewhere ~\cite{pre:parmananda}.
The advantage of using  this method in comparison to  the power
norm ($C_0(0)$)~\cite{pre:parmananda} lies in
the fact that AMD remains independent of the lag between the
measured floating potential and the applied periodic
square pulse. This is of relevance in  our experimental
system, where  invariably there exists  a lag, at times
varying in time due to the parameter drifts.
For the CR experiments it was occasionally observed
that while with an  initial increase in noise amplitude (D)
NV reaches  a minimum, the subsequent rise of
NV for even higher amplitudes of noise was suppressed.
This leads to the modification of the unimodal
profile, a signature of the CR phenomenon.
A possible explanation for this suppression is that
by virtue of the superimposed high frequency
noise (bandwidth 500 KHz) and fast responding
internal plasma dynamics, the system has the
capability of exciting high frequency regular modes
within  the ion plasma frequency (105 kHz). This in
turn  leads to the persistence of low NV values.
Finally, in Refs~\cite{prl:Lin I,pop:dinklage} both
the destructive and constructive role of noise (CR only)
have  been reported  for glow discharge and magnetized rf
discharge plasma systems respectively. However, both
these experiments were carried out in the vicinity of the
Hopf bifurcation.  In contrast, for the present work  we studied both
stochastic (SR) and coherence resonance (CR) in
the neighborhood of the homoclinic bifurcation.

\appendix
\section{Calculation of Absolute Mean Difference(AMD)}

Regularity in the stochastic resonance has been done by calculating cross-correlation ($C_o=|<[(x_1-<x_1>)(x_2-<x_2>)]>|$). But in this case this is not suitable, because, as we also measuring floating potential at different location in side the plasm there is always a lag between periodic signal that is applied in the plasma and output. This lag also varies with time because the plasma conditions are changing continuously with time. Therefore cross-correlation between output and input  signal gives wrong estimation. So we have proposed a statistics which will be independent of lag and defined as follows:

1. First calculate mean inter-peak distance ($\delta$) of the periodic signal.

2. Calculate inter-peak distances ($t_p$) of the output signal.

3. Calculate ($(t_p-\delta)/\delta$)

4. Take absolute.

5. This will perfectly describe the regularity for stochatic resonance for periodic subthreshold signa.

and this may be called Absolute Mean Difference (AMD).

$AMD=|<(\frac{t_p}{\delta}-1)>|$
where $t_p$, $\delta$ are the output signal inter-peak distance and mean peak distance of the subthreshold periodic signal respectively.


\begin{thebibliography}{100.}

\bibitem{JPhysA:benzi} {Roberto Benzi, Alfonso Sutera, and Angelo Vulpiani, J. Phys. A: Math Gen. \textbf{14}, L453-L457 (1981)}.
\bibitem{prl:gang} {Hu Gang, T. Ditzinger, C. Z. Ning, and H. Haken, Phys. Rev. Lett. \textbf{71},
807 (1993)}.
\bibitem{prl:pikovsky} {Arkady S. Pikovsky and J\"{u}rgen Kurths,Phys. Rev. Lett. \textbf{78},
777 (1997)}.
\bibitem{prl:bruce} {Bruce Mc Namara, Kurt Wiesenfeld, and Rajarshi Roy, Phys. Rev. Lett. \textbf{60},
2626 (1988)}.
\bibitem{pre:parmananda} {P. Parmananda, Gerardo J. Escalera Santos, M. Rivera, and Kenneth Showalter, Phys. Rev. E \textbf{71}, 031110 (2005)}.

\bibitem{JStatPhys:moss} {F. Moss, A. Bulsara, and M. F. Shlesinger, J. Stat. Phys. \textbf{70}, 1
(1993)}.
\bibitem{prl:longtin} {A. Longtin, A. Bulsara, and F. Moss, Phys. Rev. Lett. \textbf{67}, 656
(1991)}.
\bibitem{JPhysChem:foster} { A. F\"{o}ster, M. Merget, and F. W. Schneider, J. Phys. Chem.
\textbf{100}, 4442 (1996)}.
\bibitem{JPhysChem:Amemiya} { T. Amemiya, T. Ohmori, M. Nakaiawa, and T. Yamaguchi, J.
Phys. Chem. \textbf{102}, 4537 (1998)}.
\bibitem{pre:santos} { G. J. Escalera Santos and P. Parmananda, Phys. Rev. E \textbf{65},
067203 (2002)}.
\bibitem{prl:kitajo} {Keiichi Kitajo, Daichi Nozaki, Lawrence M. Ward and Yoshiharu Yamamoto, Phys. Rev. Lett.\textbf{90}, 218103 (2003)}.

 \bibitem{prl:Lin I} {Lin I and Jeng-Mei Liu, Phys. Rev. Lett. \textbf{74}, 3161 (1995)}.
\bibitem{pop:dinklage} { A. Dinklage, C. Wilke and T. Klinger, Phys. Plasmas \textbf{6}, 2968 (1999)}.
\bibitem{prl:Giacomelli} {Giovanni Giacomelli, Massimo Giudici, Salvador Balle, and Jorge R. Tredicce, Phys. Rev. Lett.\textbf{84}, 3298 (2000)}.
\bibitem{pre:miyakawa} {Kenji Miyakawa and Hironobu Isikawa, Phys. Rev. E \textbf{66}, 046204 (2002)}.
\bibitem{pre:Santos1} {Gerardo J. Escalera Santos, M. Rivera, M. Eiswirth, and P. Parmananda, Phys. Rev. E \textbf{70}, 021103 (2004)}.
\bibitem{prl:avila} {Jhon F. Martine Avila, Hugo L. D. de S. Cavacante, and J.R. Rios Leite, Phys. Rev. Lett. \textbf{93}, 144101 (2004)}.

\bibitem{pr:duncan} {Duncan H. Looney, and Sanborn C. Brown, Phys. Rev. \textbf{93}, 965 (1954)}.
\bibitem{pr:sturrock} {P. A. Sturrock, Phys. Rev. \textbf{117}, 1426 (1960)}.
\bibitem{pop:Shokri} {B. Shokri, S. M. Khorashadizadeh, Phys. Plasmas \textbf{13}, 052116 (2006)}.
\bibitem{Physleta:jaman} {Md. Nurujjaman, and A.N.Sekar Iyengar, Phys Letts A \textbf{360}, 717 (2007)}.
\bibitem{chaos:Nurujjaman} {Md. Nurujjaman, Ramesh Narayanan, and A. N. Sekar Iyengar, CHAOS \textbf{17}, 043121 (2007)}.


\end{thebibliography}
\end{document}